\documentclass[twocolumn,showpacs,preprintnumbers,amsmath,amssymb]{revtex4}
\usepackage[dvips]{graphicx}
\usepackage{dcolumn}
\begin{document}


\title
{Pairing States of Superfluid $^3$He in Uniaxially Anisotropic Aerogel
}

\author{Kazushi Aoyama and Ryusuke Ikeda}

\affiliation{%
Department of Physics, Graduate School of Science, Kyoto University, Kyoto 606-8502, Japan
}

\date{\today}

\begin{abstract}
Stable pairing states of superfluid $^3$He in aerogel are examined in the case with a {\it global} uniaxial anisotropy which may be created by applying an uniaxial stress to the aerogel. Due to such a global anisotropy, the stability region of an ABM pairing state becomes wider. In an uniaxially {\it stretched} aerogel, the {\it pure} polar pairing state with a horizontal line node is predicted to occur, as a 3d superfluid phase, over a measurable width just below the superfluid transition at $T_c({\rm P})$. A possible relevance of the present results to the case with no global anisotropy is also discussed. 
\end{abstract}


\maketitle

In superconductivity and superfluidity, an intrinsic anisotropy has a profound 
effect on the resulting pairing state. For instance, one main origin of the d$_{x^2-y^2}$ pairing state in high $T_c$ cuprates \cite{highTc} is the four-fold symmetry of the Fermi surface resulting from the crystalline anisotropy. Due to such an intrinsic and global anisotropy, a specific pairing symmetry with the highest temperature of Cooper instability is realized in the case of anisotropic superconductivity with no degeneracy between different pairing states. In contrast, the bulk liquid $^3$He has an isotropic Fermi surface, and hence, there is a degeneracy in the transition point between different pairing states at least when the fluctuation is neglected \cite{VW,AI}. Recently, possible pairing states of superfluid $^3$He in {\it globally isotropic} aerogels were examined by comparing the free energy in the Ginzburg-Landau (GL) region between different states, and it was found \cite{AI} that, after averaging over the quenched disorder brought by the aerogel structure, the pairing states to be realized are unaffected by the {\it locally} anisotropic scattering events due to the aerogel. However, situation may change if, as in the superconducting case mentioned above, a global anisotropy \cite{HS,Florida} is introduced externally in the aerogel 
structure. 

Here, we report on our results of theoretical phase diagrams of superfluid $^3$He in an aerogel with {\it global} anisotropy. Throughout this paper, we assume that such an anisotropy may be induced in quasiparticle scattering events by applying an uniaxial stress \cite{HS,Florida} to aerogels. Based on a conventional model \cite{Thuneberg} of effects of aerogel on $^3$He, the anisotropy can be incorporated in a momentum dependence of the {\it random-averaged} quasiparticle scattering amplitude. When the aerogel is uniaxially compressed, the 2d-like pairing state is favored, and the region of the ABM pairing state becomes wider. In the stretched case, a 1d-like pairing state is favored at least near $T_c$, and actually the {\it pure} polar pairing state should be realized accompanied by a 2nd order transition to a deformed ABM state at a lower temperature. This is a rare situation in which a new pairing state is expected to occur as a 3d superfluid phase of $^3$He. This research was preliminarily reported elsewhere \cite{RIf}. 

Just as in Ref.\cite{AI}, we start from the BCS hamiltonian with an impurity potential term 
\begin{equation}
{\cal H}_{\rm imp}=\int d^3r \sum_\sigma \psi^\dagger_\sigma({\bf r})u({\bf r})\psi_\sigma({\bf r}),\\
\end{equation}
where $u({\bf r})$ denotes an impurity potential for quasiparticles brought by aerogel structures. As argued in Ref.\cite{AI}, the scattering amplitude $|u_{\bf k}|^2$ in an aerogel, where $u_{\bf k}$ is the Fourier transform of $u({\bf r})$, should have a highly anisotropic and complicated momentum dependence, reflecting {\it local} anisotropy in aerogels. However, as far as the aerogel is globally isotropic, the local anisotropy is not reflected in averaged quantities such as ${\overline {|u_{\bf k}|^2}}$, where the overbar denotes the average over the impurity configuration due to the aerogel. On the other hand, in a globally anisotropic aerogel, ${\overline {|u_{\bf k}|^2}}$ remains anisotropic. We invoke the model \cite{Sauls} 
\begin{equation}
{\overline {|u_{\bf k}|^2}} = A (1+\delta_u({\hat {\bf k}}\cdot\hat{z})^2), 
\end{equation}
where $\delta_u$ is a small parameter measuring the global anisotropy, 
${\hat {\bf k}} = {\bf k}/k_{\rm F}$, and ${\hat z}$ denotes the direction of a uniaxial deformation. 
Although the factor $A$ is a function of ${\bf k}^2$, such an isotropic 
${\bf k}$ dependence induces no difference between various pairing states, and hereafter, $A$ will be treated as a constant factor. Then, within the Born approximation, the quasiparticle Green's function $G_\varepsilon({\bf p}) = (i \varepsilon - \xi_{\bf p} - \Sigma_{\bf p})^{-1}$ satisfies 
\begin{equation}
\Sigma_{\bf p} = \int_{\bf p'} \, \, \frac{\overline {|{u_{{\bf p}-{\bf p'}}}|^2}}{i\varepsilon-\xi_{\bf p'}-\Sigma_{{\bf p}'}}, 
\end{equation}
where $\varepsilon$ is the fermionic Matsubara frequency, and $\quad \int_{\bf p'}=\int d^3p'/(2\pi)^3$. 
Below, we will focus on the situations satisfying $2 \pi \tau T \gg 1$, where $\tau^{-1} = 2 \pi N(0) \langle {\overline {|u_{\bf k}|^2}} \rangle_{\hat {\bf k}}$ is the relaxation rate. Taking $\Sigma_{\bf p}$ to be purely imaginary and performing the ${\bf p'}$-integral close to the Fermi surface, we obtain $\Sigma_{\bf p} = -i \, \eta_{\bf p} \, {\rm sgn} \, \varepsilon$, where 
\begin{equation}
\eta_{\bf p} = \frac{1}{2\tau}(1+{\tilde \delta}_u(\hat{p}\cdot\hat{z})^2), 
\end{equation}
and ${\tilde \delta}_u = {\delta_u} (1 + \delta_u/3)^{-1}$. 
The neglect of the real part, Re($\Sigma_{\bf p}$), of $\Sigma_{\bf p}$ is safely valid as far as $2 \pi \tau T \gg 1$: The ${\hat {\bf p}}$-dependence in Re($\Sigma_{\bf p}$) may be absorbed into an anisotropy of density of states (DOS), while such a correction to DOS is scaled by $E_{\rm F}^{-1}$. Hence, the neglected correction to Re($\Sigma$) is of the order $\delta_u/(E_{\rm F} \tau)$, which is smaller than the magnitude of particle-hole asymmetry $T_c/E_{\rm F}$. In this manner, focusing on the imaginary part of $\Sigma$ is justified. Then, $G_\varepsilon({\bf p})$ becomes 
\begin{equation}
G_\varepsilon({\bf p})=(i(|\varepsilon| + \eta_{\bf p}){\rm sgn} \, \varepsilon - \xi_{\bf p})^{-1}. 
\end{equation}

Note that, when $\delta_u > 0$ ($< 0$), the mean free path of normal 
quasiparticles running along the $z$-direction is shorter (longer). Thus, the case with $\delta_u > 0$ ($< 0$) corresponds to the uniaxially compressed (stretched) case. 

Besides the self energy part, the vertex part in the particle-particle channel is also affected by the impurity scattering. If neglecting spatial variations of the pair-field $A_{\mu,i}$, the bare vertex ${\hat p}_i$ is replaced by $\Gamma_i(\varepsilon, {\bf p})$, where  
\begin{equation}\label{eq:ppv}
\Gamma_i(\varepsilon,{\bf p})=\hat{p}_i+\int_{\bf p'}\Gamma_i(\varepsilon,{\bf p'}) \, |G_{\varepsilon}({\bf p'})|^2 \, {\overline {|u_{\bf p-p'}|^2}}.
\end{equation}
The solution of eq.(\ref{eq:ppv}) takes the form $\Gamma_i(\varepsilon,{\bf p})=\hat{p}_i+(\hat{p}\cdot\hat{z})\overline{\Gamma}_\varepsilon\hat{z}_i$, where 
\begin{eqnarray}
\overline{\Gamma}_\varepsilon 
&=&-1+ \Bigg[ \displaystyle{1-2\sum_{k\ge 1}\frac{1}{2k+1}\bigg(\frac{-{\tilde \delta}_u\tau^{-1}}{2|\varepsilon|+\tau^{-1}}\bigg)^k} \Bigg]^{-1}.  
\end{eqnarray}
The above expressions will be used to derive a GL hamiltonian per volume $h_{\rm GL}$ in the anisotropic case. Up to O($\delta_u$), its quadratic term is expressed by 
\begin{eqnarray}\label{eq:quadr}
&h_{\rm GL}^{(2)}& = A_{\mu,i}^\ast A_{\mu,j}\Bigg[\frac{N(0)}{3}\Big(\ln\frac{T}{T_{c0}}+T\displaystyle{\sum_\varepsilon\frac{\pi}{|\varepsilon|}}\Big)\delta_{i,j} \nonumber \\
&-& T\sum_\varepsilon\int_{\bf p}\hat{p}_i\Gamma_j(\varepsilon,{\bf p})G_{\varepsilon}({\bf p})G_{-\varepsilon}(-{\bf p})\Bigg]\nonumber\\
&\simeq& \frac{N(0)}{3}\Bigg[\ln\frac{T}{T_{c0}}+\psi\big(\frac{1}{2}+\frac{1}{4\pi T\tau}\big)-\psi\big(\frac{1}{2}\big)\nonumber\\
&+& \frac{\delta_u}{4\pi T\tau}\frac{1}{5}\psi^{(1)}\big(\frac{1}{2}+\frac{1}{4\pi T\tau}\big) \Bigg] A_{\mu,i}^{\ast}A_{\mu,i} \nonumber \\
&+& \frac{N(0)}{3} \frac{\delta_u}{4\pi T\tau}\frac{16}{15}\psi^{(1)}\big(\frac{1}{2}+\frac{1}{4\pi T\tau}\big) \, A_{\mu,z}^{\ast}A_{\mu,z}, 
\end{eqnarray}
where $\psi(z)$ is the di-gamma function, and $A_{\mu,i} \equiv |\Delta(T)| a_{\mu,i}$ with $a_{\mu,i}(a_{\mu,i})^* = 1$. The anisotropic term due to the vertex correction contributes to eq.(\ref{eq:quadr}) with the same sign as that due to $\Sigma_{\bf p}$. 

The quartic term $h_{\rm GL}^{(4)}$ is also derived in a similar 
manner and, up to O($\delta_u$), takes the form 
\begin{eqnarray}
&h_{\rm GL}^{(4)}& \, = \, \beta_1 |A_{\mu,i}A_{\mu,i}|^2+\beta_2 (A_{\mu,i}^{\ast}A_{\mu,i})^2 
+ \beta_3 A_{\mu,i}^{\ast}A_{\nu,i}^{\ast}A_{\mu,j}A_{\nu,j} \nonumber \\ 
&+& \beta_4 A_{\mu,i}^{\ast}A_{\nu,i}A_{\nu,j}^{\ast}A_{\mu,j} + \beta_5 A_{\mu,i}^{\ast}A_{\nu,i}A_{\nu,j}A_{\mu,j}^{\ast} \nonumber \\
&+& [ \, ( \, \beta^{(1)}_1  A_{\mu,i}A_{\mu,i}A_{\nu,z}^{\ast}
A_{\nu,z}^{\ast} 
+ \beta^{(1)}_2 A_{\mu,i}^{\ast}A_{\mu,i}A_{\nu,z}^{\ast}A_{\nu,z} \nonumber \\
&+& \beta^{(1)}_3 A_{\mu,i}A_{\nu,i}A_{\mu,z}^{\ast}A_{\nu,z}^{\ast} 
 + \beta^{(1)}_4 A_{\mu,i}^{\ast}A_{\nu,i}A_{\nu,z}^{\ast}A_{\mu,z} 
 \nonumber \\ 
&+& \beta^{(1)}_5 A_{\mu,i}^{\ast}A_{\nu,i}A_{\mu,z}^{\ast}A_{\nu,z} \, ) + {\rm c.c.} \, ]. 
\end{eqnarray}
Each of the coefficients $\beta_i$ is the sum of a weak coupling contribution $\beta_i^{(0)}$ and a strong coupling one $\delta\beta_i$. Regarding $\delta\beta_i$, their expressions with $\delta_u=0$ derived in Ref.\cite{AI} will be used hereafter. This approximation should not affect calculation results except for extremely large $|\delta_u|$ values. 
The coefficients $\beta^{(0)}_i$ and $\beta^{(1)}_i$
are given by 
\begin{eqnarray}
\beta^{(0)}_3 &=& -2\beta^{(0)}_1 = - \, \frac{\beta_0(T)}{7\zeta(3)}\Big[\psi^{(2)}\big(\frac{1}{2}+\frac{1}{4\pi T\tau}\big) \nonumber \\
&+&\frac{\delta_u}{4\pi T\tau}\frac{1}{7} \psi^{(3)}\big(\frac{1}{2} +\frac{1}{4\pi T\tau}\big) \Big], \nonumber \\
\beta^{(0)}_2&=&\beta^{(0)}_4=-\beta^{(0)}_5 \nonumber\\
&=& \beta^{(0)}_3-\frac{1}{4\pi T\tau}\frac{\beta_0(T)}{7\zeta(3)}\Big[\big(\frac{5}{18}+\frac{\delta_u}{54}\big)\psi^{(3)}\big(\frac{1}{2}+\frac{1}{4\pi T\tau}\big) \nonumber \\
&+&\frac{\delta_u}{4\pi T\tau} \frac{1}{18} \psi^{(4)}\big(\frac{1}{2}+\frac{1}{4\pi T\tau}\big)\Big], \nonumber\\
\beta^{(1)}_3&=&-2\beta^{(1)}_1 = - \, \frac{\delta_u}{4\pi T\tau}\frac{\beta_0(T)}{7\zeta(3)} \frac{46}{63}\psi^{(3)}\big(\frac{1}{2} 
+ \frac{1}{4\pi T\tau}\big), \nonumber \\
\beta^{(1)}_2&=&\beta^{(1)}_4=-\beta^{(1)}_5\nonumber\\
&=& \beta^{(1)}_3-\frac{\delta_u}{4\pi T\tau}\frac{\beta_0(T)}{7\zeta(3)}\Big[\frac{1}{9}\psi^{(3)}\big(\frac{1}{2}+\frac{1}{4\pi T\tau}\big) \nonumber \\
&+& \frac{1}{4\pi T\tau} \frac{4}{27} \psi^{(4)}\big(\frac{1}{2}+\frac{1}{4\pi T\tau}\big)\Big] 
\end{eqnarray}
up to O($\delta_u$), where 
$\beta_0(T)=7 \zeta(3) N(0)/{(240 \pi^2 T^2)}$. 

As seen in eq.(\ref{eq:quadr}), the inclusion of a global anisotropy induces a splitting of the Cooper instability point between different pairing states. Since $\psi^{(1)}(y) > 0$ ($y > 0$), an uniaxial compression with positive $\delta_u$ makes the instability point of 2d-like pairing states with vanishing $a_{\mu,z}$ higher, implying that such a state must be realized just below $T_c(P)$. Situation is similar to $^3$He thin films, and thus, this 2d-like state should be the ABM state. In the same manner, in the uniaxially stretched case with negative $\delta_u$, the 1d-like polar pairing state with $a_{\mu,i} = {\hat d}_{\mu} \delta_{i,z}$ tends to occur just below $T_c$. However, it is unclear at this stage whether or not the state just below $T_c$ may be a mixture of the ABM and polar pairing states \cite{Ho} so that the pure polar symmetry obtained at $T_c$ crosses over upon cooling to the ABM one with no transition. 

It should be noted that, in the disordered case, additional terms are induced in the GL hamiltonian by the impurity scattering and its local anisotropy. When the global anisotropy is absent, a combination of one of such terms and the gradient term, which was not represented in eq.(8), leads to destruction of superfluid long-ranged order (LRO) in the ABM state \cite{AI,Volovik}. However, the contribution to the free energy of the additional term is well described simply by incorporating a disorder-induced shift of $T_c$ into the {\it mean field} condensation energy $E_c$ \cite{AI}. In determining phase diagrams below, we have followed this finding \cite{AI} and, for brevity, have used the disorder-induced $T_c$-shift with $\delta_u=0$. The latter procedure does not affect our quantitative results unless {\it all} transitions between different pairing symmetries occur in the close vicinity of $T_c$. 
\begin{figure}[t]
\scalebox{0.55}[0.55]{\includegraphics{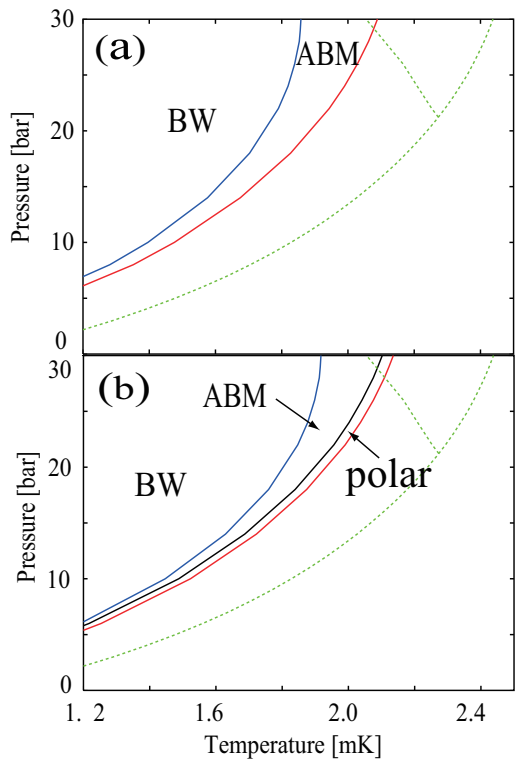}}
\caption{Obtained $P$-$T$ phase diagrams in an uniaxially-compressed case (a) with $\delta_u = + 0.04$ and a stretched case (b) with $\delta_u = - 0.07$. For both figures, we have used $(2 \pi \tau)^{-1} = 0.137$ (mK). The solid curves are transition lines in aerogel, while the thin dotted curves are those of bulk liquid. The ABM state in (a) is a genuine superfluid with LRO, while in (b) it is a superfluid {\it glass} \cite{Volovik,AI}.}
\label{fig.1}
\end{figure}
\begin{figure}[t]
\scalebox{0.5}[0.5]{\includegraphics{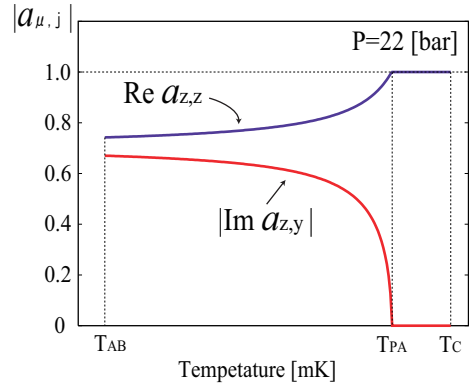}}
\caption{Variation of $a_{\mu,i}$ at $P=22$ (bar) around the 2nd order transition at $T_{\rm PA}$ between the polar and the deformed ABM states with $|{\rm Re}a_{z,z}| \neq |{\rm Im}a_{z,y}|$. The same material parameters were used as in Fig.1 (b). A 1st order transition to the deformed BW state occurs at $T_{\rm AB}$.} 
\label{fig.2}
\end{figure}

We have numerically examined transitions between different pairing states through $E_c$ by taking account of $T_c$-shifts of different origins mentioned above. Since fully including 18 real components of the pair 
field $a_{\mu, i}$ is cumbersome, nonvanishing four 
components, ${\rm Re} \, a_{\mu, \mu}$ and ${\rm Im} \, a_{z,y}$, were kept in calculations so that the familiar ABM. BW, planar, and polar 
states are taken into account. For instance, in the uniaxially stretched case, the route through 
which only ${\rm Re}a_{z,z}$ and ${\rm Im}a_{z,y}$ remain nonzero is found to be the most favorable upon cooling from the Cooper instability of the polar state (see Fig.2). Typical examples of the resulting phase diagrams for $\delta_u > 0$ (compressed case) and $\delta_u < 0$ (stretched case) are given in Fig.1 (a) and 1 (b), respectively, where the GL hamiltonian valid up to O($\delta_u^2$) was used. In both of stretched and compressed cases, there is no polycritical pressure (PCP), i.e., there is a nonvanishing $P$-range of the ABM state even in low $T$ limit. For the used $\tau$-value, there is no {\it mean field} stability region in $P < 30$ (bar) of the ABM state in the isotropic ($\delta_u=0$) case. It implies that the main origin inducing the ABM state in the figures is the $T_c$-shift and not the strong coupling effect. Further, we have verified that the planar pairing state cannot overcome the ABM one in free energy at any $P$ and $T$ and is not realized as a pairing state in equilibrium. 

In the ABM state in the compressed case, the direction of gap nodes is pinned on average along ${\hat z}$ just like in thin films where the ABM or planar state has a wider stability region \cite{HS}. An origin of the remarkably wide ABM region in Fig.1 (a) can be attributed to the similarity to the thin film case. Further, due to this pinning effect of ${\bf l}$-vector, this ABM state has a true superfluid LRO in contrast to the quasi LRO in the isotropic case \cite{AI}. On the other hand, in the stretched case (or equivalently, the case compressed in the cylindrically symmetric manner), the direction of gap nodes in the deformed ABM state (see Fig.2) is {\it spontaneously} chosen within the $x$-$y$ plane. Situation is similar to the bulk $^3$He in a uniform magnetic field which also favors the ABM state. Thus, the ABM stability region becomes wider even in the stretched case, although this state is a superfluid {\it glass} with no genuine superfluid  LRO \cite{AI}. Further, in both cases, the BW state with no gap nodes is deformed by the anisotropy (i.e, $a_{x,x}=a_{y,y} 
\neq a_{z,z}$) \cite{comlro}. 

An intriguing result in the uniaxially-stretched case is the appearance of the pure polar pairing state, with a horizontal line of gap nodes in $x$-$y$ plane, just below $T_c(P)$-line. It appears that the temperature width over which the polar state is stable will be observable experimentally. Needless to say, this temperature width is, as well as that of the ABM state, extended with increasing $|\delta_u|$. This polar state is not a mixture with other pairing states and, as in Fig.2, shows a 2nd order transition to a deformed ABM state with point nodes in the $x$-$y$ plane upon cooling. A $T_c$-shift dependent on the pairing states resulting from eq.(8) is essential to obtaining the {\it pure} polar state. Since the direction ${\hat {\bf P}}$ of gap-maximum is pinned by ${\hat z}$, this polar state has a true superfluid LRO. 

\begin{figure}[t]
\scalebox{1.2}[1.2]{\includegraphics{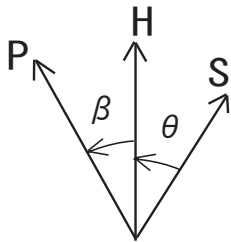}}
\caption{Initial configuration 
to be taken 
in pulsed NMR experiments in the polar state. The angles $\theta$ and $\beta$ are the tipping angle in ${\bf H}$ - ${\hat {\bf P}}$ plane and $\cos^{-1}({\hat {\bf P}} \cdot {\bf H}/H)$, respectively.} 
\label{fig.3}
\end{figure}
As a measure useful in detecting the polar state with the order parameter $A_{\mu,i} = \Delta {\hat d}_\mu \, {\hat P}_i$, a pulsed NMR frequency shift in the polar state will be considered. Here, ${\hat {\bf P}}$ points on averages to the stretched direction ${\hat z}$. Fluctuations of ${\hat {\bf P}}$ are assumed to be negligibly small. As usual, effects of the dipole energy on the spin dynamics can be examined in terms of Leggett's equations \cite{Leggett} as far as the initial configuration is in equilibrium. In the pure polar state, the so-called dipole torque ${\bf R}_D$ \cite{VW} is given by 
\begin{equation}
{\bf R}_D= - \, \displaystyle{\frac{12g_D}{5}}(\hat{\bf d}\cdot\hat{\bf P})(\hat{\bf d}\times \hat{\bf P}),
\end{equation}
where $g_D$ is a dipole energy strength \cite{VW}, and hence, ${\hat {\bf P}} \perp {\hat {\bf d}}$ and ${\bf H} \perp {\hat {\bf d}}$ are to be satisfied in equilibrium. Then, a frequency shift $\Delta \omega$ of the free-induction signal due to ${\bf R}_D$ occuring by tipping the magnetization by an angle $\theta$ is given by 
\begin{equation}\label{eq:delome}
\Delta\omega = \frac{\Omega^2_L}{2\omega}\big[(3\cos^2\beta-1)\cos\theta+\frac{\sin^2\beta}{2}(1+\cos\theta)\big], 
\end{equation}
which depends on the angle $\beta$ spanning ${\hat {\bf P}}$ and ${\bf H}$. 

Finally, we note that the present result may also be relevant to liquid $^3$He in globally {\it isotropic} aerogels \cite{AI} if the correlation length $\xi_a$ of the {\it local} anisotropy \cite{HS} is much longer than the superfluid coherence length $\xi_0$. Since, as mentioned earlier, the free energy of each pairing state is roughly determined by the condensation energy $E_c$ even at length scales of the order of $\xi_0$, the results induced by the global anisotropy, such as the wider ABM region \cite{Florida,Sauls} and an occurrence of the polar pairing state near $T_c$, may be valid in the globally isotropic aerogel under the condition $\xi_a \gg \xi_0$ which may be satisfied at higher pressures. Then, 
a polar {\it glass} phase with spatially random ${\hat {\bf P}}$ over long distances might be realized in the equal-spin pairing region near $T_c$. In contrast, at lower pressures with longer $\xi_0$-values, the anisotropy-induced $T_c$-shift will be negligible, and the approach \cite{AI} modelling the {\it local} anisotropy as a random field becomes appropriate. Then, a well-defined PCP is expected at a nonzero temperature. 

In conclusion, by introducing a global anisotropy in aerogels, the region of A-like phase should be extended if this phase has the ABM pairing state. A recent measurement has shown an extention of the A-like phase region, even on warming, due to an uniaxial compression \cite{Halperin}. In aerogels deformed via a uniaxial stretch or a cylindrically symmetric compression, an appearance of the pure polar pairing state near $T_c$ and a wider region of ABM superfluid glass \cite{AI} are expected. We hope a measurement for the stretched case to be 
performed. 

One of authors (R.I.) is grateful to O. Ishikawa, Y. Lee, and J. A. Sauls for useful discussions. The present work is supported by the Grant-in-Aid for the 21st Century COE "Center for Diversity and Universality in Physics" from the Ministry of Education, Culture, Sports, Science, and Technology (MEXT) of Japan.


\begin{thebibliography}{99}
\bibitem{comlro} This deformed BW state is also expected to have no genuine LRO, because the "anisotropy" $a_{x,x}/a_{z,z}$ is $T$-dependent and thus, does not become compatible with the anisotropy of the quenched disorder itself.
\end{thebibliography}
\end{document}